\documentclass[a4paper,UKenglish,cleveref, autoref, thm-restate]{lipics-v2021}
\hideLIPIcs 
\nolinenumbers
\usepackage{hyperref}
\usepackage{tabularx}
\usepackage{multicol}
\usepackage{caption}
\usepackage{subcaption}
\usepackage{rotating}
\usepackage{algorithm}
\usepackage[noend]{algorithmic}
\newcommand\algorithmicprocedure{\textbf{procedure}}
\newcommand{\algorithmicendprocedure}{\algorithmicend\ \algorithmicprocedure}

\makeatletter
\newcommand\PROCEDURE[3][default]{%
  \ALC@it
  \algorithmicprocedure\ \textsc{#2}(#3)%
  \ALC@com{#1}%
  \begin{ALC@prc}%
}
\newcommand\ENDPROCEDURE{%
  \end{ALC@prc}%
  \ifthenelse{\boolean{ALC@noend}}{}{%
    \ALC@it\algorithmicendprocedure
  }%
}
\newenvironment{ALC@prc}{\begin{ALC@g}}{\end{ALC@g}}
\makeatother

\usepackage{rotating}
\usepackage{multirow}
\usepackage{numprint}
\usepackage{xspace}
\usepackage{wrapfig}

\usepackage{mathtools}

\usepackage{amssymb}

\newtheorem{reduction}{Reduction}[section]

\def\HypergraphEdges{E}

\usepackage{pgf}
\usepackage{pgfplots}
\usepackage{iftex}
\usepackage{ifxetex,ifluatex}
\newcommand{\NP}{\ensuremath{\mathbf{NP}}}

\newif\ifunicode
\ifxetex\unicodetrue\fi
\ifluatex\unicodetrue\fi
\ifluatex
\newcommand{\displaypgf}[2]{\includegraphics{#1/#2.pdf}}
\else 
\newcommand{\displaypgf}[2]{\includegraphics{#1/#2.pdf}}
\fi
\ifPDFTeX
\else

\ifunicode

  \usepackage{newunicodechar}
  \newcommand{\DeclareUnicodeCharacter}[2]{%
    \begingroup\lccode`|=\string"#1\relax
    \lowercase{\endgroup\newunicodechar{|}}{#2}%
  }
\else
\fi
\fi
\DeclareUnicodeCharacter{2212}{−}
\usepackage{tikz}
\usepgfplotslibrary{groupplots,dateplot}

\usetikzlibrary{patterns,shapes.arrows}
\pgfplotsset{compat=newest}

\tikzset{semithick/.style={line width =1.5pt}}
\tikzset{font={\small}}

\def\stripzero#1{\expandafter\stripzerohelp#1}
\def\stripzerohelp#1{\ifx 0#1\expandafter\stripzerohelp\else#1\fi}

\newcommand{\nodesOf}[1]{#1}
\newcommand{\edgesOf}[1]{E(#1)}

\def\MdN{\ensuremath{\mathbb{N}}}

\newcommand{\Return}   {{\bf return\ }}

\newcommand{\ie}{i.\,e.,\xspace}

\newcommand{\etal}{et~al.\xspace}

\def\comment#1{}

\def\withcomments{
	\newcounter{mycommentcounter}
	\def\comment##1{\refstepcounter{mycommentcounter}%
		\ifhmode%
		\unskip%
		{\dimen1=\baselineskip \divide\dimen1 by 2 %
			\raise\dimen1\llap{\tiny\bfseries \textcolor{red}{-\themycommentcounter-}}}\fi%
		\marginpar[{\renewcommand{\baselinestretch}{0.8}%
			\hspace*{3em}\begin{minipage}{5em}\footnotesize [\themycommentcounter]: \raggedright ##1\end{minipage}}]{\renewcommand{\baselinestretch}{0.8}%
			\begin{minipage}{5em}\footnotesize [\themycommentcounter]: \raggedright ##1\end{minipage}}}
}

\usepackage{graphics}
\usepackage{todonotes}
\usepackage{booktabs}
\usepackage{csquotes}
\usepackage[normalem]{ulem}

\title{\Large Engineering Hypergraph $b$-Matching Algorithms}


\date{}

\author{Ernestine Großmann}{Heidelberg University, Germany }{E.Grossmann@informatik.uni-heidelberg.de}{https://orcid.org/0000-0002-9678-0253}{}%
\author{Felix Joos}{Heidelberg University, Germany }{joos@informatik.uni-heidelberg.de}{https://orcid.org/0000-0002-8539-9641}{}%
\author{Henrik Reinstädtler}{Heidelberg University, Germany }{henrik.reinstaedtler@informatik.uni-heidelberg.de}{https://orcid.org/0009-0003-4245-0966}{}%
\author{Christian Schulz}{Heidelberg University, Germany }{christian.schulz@informatik.uni-heidelberg.de}{https://orcid.org/0000-0002-2823-3506}{}

\acknowledgements{We acknowledge support by DFG grant SCHU 2567/8-1 and by the state of Baden-Württemberg through bwHPC.}
\authorrunning{E. Großmann, F. Joos, H. Reinstädtler and C. Schulz} 
\Copyright{Ernestine Großmann, Felix Joos, Henrik Reinstädtler and Christian Schulz}

\begin{document}

\maketitle

\begin{abstract} \small%
    Recently, researchers have extended the concept of matchings to the more general problem of finding $b$-matchings in hypergraphs broadening the scope of potential applications and challenges. 
    The concept of $b$-matchings, where $b$ is a function that assigns positive integers to the vertices of the graph, is a natural extension of matchings in graphs, where each vertex $v$ is allowed to be matched to up to $b(v)$ edges, rather than just one.
    The weighted $b$-matching problem then seeks to select a subset of the hyperedges that fulfills the constraint and maximizes the weight. 
    
    In this work, we engineer novel algorithms for this generalized problem.
    More precisely, we introduce exact data reductions for the problem as well as a novel greedy initial solution and local search algorithms. These data reductions allow us to significantly shrink the input size. This is done by  either determining if a hyperedge is guaranteed to be in an optimum $b$-matching and thus can be added to our solution or if it can be safely ignored.  
    Our iterated local search algorithm provides a framework for finding suitable improvement swaps of edges.
    Experiments on a wide range of real-world hypergraphs show that our new set of data reductions are highly practical, and our initial solutions are competitive for graphs and hypergraphs as well.
    
\end{abstract}

\clearpage
\setcounter{page}{1}%

\section{Introduction}
\label{sec:introduction}
Graph theory has long been a crucial discipline in the world of mathematics and computer science, providing insights into numerous complex problems. One of the most well-known problems in graph theory is the matching problem.
A matching in a graph is a set of pairwise vertex-disjoint edges.
Computing (these) matchings in a graph is a ubiquitous combinatorial 
problem that has a myriad of applications in various fields~\cite{Halappanavar09algorithms}.
By now, maximum weight/cardinality matchings in graphs in the internal-memory model have been extensively studied, leading to various breakthroughs. However, finding maximum (weight) matchings in graphs in the internal-memory model is only the tip of the iceberg.
Recently, researchers have extended the concept of matchings to the more general problem of finding $b$-matchings in hypergraphs broadening the scope of potential applications and challenges. 
A hypergraph is a natural graph extension in which edges can have more than two endpoints and thus model more complex relationships.
 For example, online hypergraph matching can be used to model auctions of advertisement campaigns~\cite{korula2009algorithms}.
Moreover, the concept of $b$-matchings (with $b: V \to \MdN$) is a natural extension of matchings, where each vertex $v$ is allowed to be matched to up to $b(v)$ edges.%
The weighted $b$-matching problem then seeks to select a subset of the hyperedges that fulfill the constraint and maximize~the~weight. 

Many applications require the computation of a matching $\mathcal{M}$ with certain properties, like being maximal (no edge can be added  without violating the matching property), having maximum cardinality, or having maximum total weight $\sum_{e\in \mathcal{M}}\omega(e)$.
For example, in multi-level (hyper)graph partitioning, the problem of coarsening a (hyper)graph without losing the characteristics of the original (hyper)graph in multi-level decomposition algorithms can be solved by computing a hypergraph matching problem~\cite{schlag2016k}.
Similarly, hypergraph $b$-matching plays a critical role in agglomerative hypergraph clustering~\cite{DBLP:reference/crc/PapaM07}, where hyperedges are evaluated based on the likelihood of merging adjacent clusters. In this context, the function $b$ assigned to the vertices can serve as a mechanism to regulate the pace of agglomeration. Other important example applications include
 allocating resources to machines or auctioning goods~\cite{cygan2013sell}, ride-sharing \cite{pavone2022online} and load balancing~\cite{DBLP:journals/access/HouMKZ22}. %
\setcounter{page}{1}

Currently, however, researchers have only developed approximation algorithms \cite{doi:10.1137/1.9781611977554.ch41} for the weighted case and practical implementations of heuristics to tackle the hypergraph matching problem is only limited to special classes of hypergraphs without weight~\cite{dufosse2019effective}.  
One powerful technique for tackling \NP{}-hard graph problems is to use \emph{data reduction rules}, which remove or contract local (hyper)graph structures to reduce the input instance to an equivalent but smaller instance. Originally developed as a tool for parameterized algorithms~\cite{cygan2015parameterized}, data reduction rules have been effective in practice for computing a (weighted) maximum independent set~\cite{chang2017computing,lamm2017finding,strash2016power} / minimum vertex cover~\cite{akiba-tcs-2016}, maximum clique~\cite{chang2020,verma2015}, and maximum $k$-plex~\cite{conte2021meta,jiang2021new}, as well as solving graph coloring~\cite{verma2015,lin2017reduction} and clique cover problems~\cite{gramm2009data,strash2022effective}, among others~\cite{Abu-Khzam2022}. However, recent work has only scratched the surface for \emph{weighted} problems, with some examples being \cite{lamm2019exactly,gellner2021boosting,gu2021towards,wang2020reduction}.
\vspace*{-0.25cm}
\subparagraph*{Our Results.} In this work, we devise and engineer data reduction rules and new greedy initial solution algorithms for the weighted hypergraph $b$-matching problem in general hypergraphs. %
 Furthermore, we present a local search for this problem.
 While our main focus is on the most general weighted hypergraph $b$-matching problem, we also compare our greedy initial solutions on graphs against solvers that are restricted to graphs.  
Our experiments show that we are able to obtain better initial
solutions by greedy heuristics of up to $10$ \%, a speedup of $6.85$ for exactly solving hypergraph $b$-matching, and quality
improvements of up to 30 \% by our local search algorithm for the $1$-matching~case.

\section{Preliminaries}
\label{sec:preliminaries}
\subsection{Basic Concepts}
\label{subsec:basic_concepts}
A \textit{weighted undirected hypergraph} $H=(V,E,\omega)$ is defined as a set of $n$ vertices $V$ and a
multiset of $m$ hyperedges $E$ with  edge weights $\omega:E \rightarrow \mathbb{R}_{>0}$, where each edge~$e\in E$ is a subset of the vertex set $V$.
We define  $\hat{\omega}(S) := \{\omega(x)\mid x\in S\}$ for some subset $S\subset E$.
We assume hyperedges to be sets rather than multisets; that is, a vertex can only be contained in a hyperedge \emph{once}, while multiple edges can contain the same set of vertices. Therefore, we write $\nodesOf{e}$ for the set of vertices of a hyperedge~$e$ and define $\lvert\nodesOf{e} \rvert$ as the edge size. The maximum edge size is denoted by $\Delta_E:=\max_{e\in E}\lvert\nodesOf{e} \rvert$.
We refer to the edges of a vertex by $\edgesOf{v}:= \{e\in E \mid v\in \nodesOf{e}\}$ and for a (multi-)set $M$ of edges we define $M(v):=\edgesOf{v}\cap M$. A vertex $v$ is \textit{incident} to an edge~$e$ if $v \in \nodesOf{e}$.
The degree of a vertex $v$ is $\lvert \edgesOf{v}\rvert$ and $\Delta_{V}:=\max_{v\in V}\lvert \edgesOf{v}\rvert$ is the maximum degree.  Two vertices $u,v$ are \emph{adjacent} if at least one edge is incident to both of them. Furthermore, two edges~$e,f$ are \emph{adjacent} if $\nodesOf{e}\cap \nodesOf{f}\neq \emptyset$. We call two edges~$e,f$ \emph{linked} if there are only  vertices of degree $2$ incident to $e$ and $f$. A set of edges $S$ in $H$ is \emph{independent} if for all distinct $f,g \in S$  $\nodesOf{f}$ and $\nodesOf{g}$ are disjoint. We define $\mathcal{N}(e):=\bigcup_{v\in \nodesOf{e}}\edgesOf{v}$ as the closed neighborhood of an edge.
We extend the weight function $\omega$ to sets naturally, that is,  $\omega(F) :=\sum_{e \in F} \omega(e)$.
Given a subset $V' \subset V$, the \emph{subhypergraph} $H_{V'}$ is defined as $H_{V'}:=(V', \{e \cap V'~|~e \in E : e \cap V' \neq \emptyset \})$.
For a set of edges $R$ of a hypergraph $H=(V,E,\omega)$ we write $H\setminus R$ short for $(V,E\setminus R,\omega)$.
If the nodes of a hypergraph can be partitioned into $d$ sets such that no edge is adjacent to two vertices in the same set, a hypergraph is called $d$-partite. If each edge has the same number $d$ of vertices, a hypergraph is called $d$-uniform.
A \textit{weighted undirected graph} $G=(V,E,\omega)$ is defined as a set of $n$ vertices $V$ and a
set of $m$ edges $E$ with  edge
weights $\omega:E \rightarrow \mathbb{R}_{>0}$.
In contrast to hypergraphs, the size of the edges is restricted to~$2$.
Throughout this paper, we use \emph{edges} in the context of hypergraphs~and~graphs.

\textbf{(Hyper)graph $b$-Matching.}
A matching $\mathcal{M} \subset E$ in a (hyper)graph is a set of (hyper)edges that are pairwise disjoint.
The \emph{cardinality} or \emph{size} of a matching is simply the cardinality of the (hyper)edge subset $\mathcal{M}$.
We call a matching \emph{maximal} if there is no (hyper)edge in $E$ that can be added to $\mathcal{M}$. 
A \textit{maximum cardinality matching} $\mathcal{M}_{c}$ is a matching that contains the largest possible
number of (hyper)edges of all matchings.
A \textit{maximum weight matching} $\mathcal{M}_{\omega}$ is a matching that maximizes $\omega(\mathcal{M}_{\omega})$ among all possible matchings.
In a \emph{perfect} matching, every vertex is incident to an edge contained in the matching. 
For a given function ${b: V \to \MdN}$, the $b$-matching problem relaxes the edge-disjointness constraint so that each vertex can be incident to $b(v)$ edges. 
We define $b(v)$ as the capacity of vertex~$v$ and denote  $\beta=\beta(b) = \max_{v \in V} b(v)$.
For $b \equiv 1$, this is equivalent to the standard matching problem.
By ${blocked(e,\mathcal{M}):= \{v\in \nodesOf{e} \mid \lvert \mathcal{M}(v)\rvert=b(v)\}}$ we refer to the set of  all vertices of an edge~$e$ where the capacity is exhausted, in other words, we cannot add further edges to the matching. Similarly, we define ${blockedEdges(e):=\bigcup_{v\in \nodesOf{e}:b(v)=1}\edgesOf{v}\setminus\{e\}}$ as the edges blocked by an edge~$e$. An edge~$e$ for which $blocked(e,\mathcal{M})=\emptyset$ is called \emph{free}. 
Finally, for a finite set $X\subset \mathbb{R}_{> 0}$ let  $\mathrm{nmax}(X,k)$ denote its $k$-th largest value if it exists, otherwise $0$. 
\vspace{-0.25cm}
\subsection{Related Work}
\label{subsec:related_work}
There is a vast amount of literature for matchings in graphs \cite{edmonds1965paths,DBLP:conf/focs/MicaliV80,gabow1990data,DBLP:conf/esa/KorenweinNNZ18,DBLP:conf/alenex/DroschinskyMT20,DBLP:conf/europar/BirnOSSS13,Preis99,drake2003simple,manne2014new,duanlineartimeweightedmatching}. We refer the reader to the respective papers for more details.
For results in data reduction, we refer the reader to the recent survey \cite{Abu-Khzam2022}.
We now cover related work closer to our main contribution, which are problem variations of the most general weighted hypergraph $b$-matching problem.

\textbf{Graph $b$-Matching.}
The $b$-matching problem can be reduced to the simple matching problem according to Gabow~\cite{gabow1983efficient} by substituting vertices, but this is impractical on large graphs.
An overview of exact approaches can be found in  M{\"u}ller-Hannemann and Schwartz~\cite{muller2000implementing}. 
Grötschel and Holland~\cite{grotschel1985solving} use the cutting plane technique  to tackle the problem.
Based on belief propagation and assuming a unique solution exists, Huang and Jebara~\cite{huang2011fast} developed an exact algorithm for the $b$-matching problem.
Mestre~\cite{mestre2006greedy} proved that the greedy algorithm is a half-approximation and generalized the \textsf{PGA} algorithm by Drake and Hougardy~\cite{drake2003simple} to achieve an $O(\beta m)$ time half-approximation. The \textsf{LD} algorithm was generalized to $b$-matching by Georgiadis and Papatriantafilou~\cite{georgiadis2013overlays} in a distributed fashion. 
Khan et al.~\cite{khanbmatching} introduced an approximation algorithm that can be executed in parallel called \textsf{bSuitor}, inspired by the results of Manne and Halappanavar~\cite{manne2014new} for normal matchings. 
Ferdous \etal \cite{DBLP:conf/acda/FerdousP0PH21} consider parallel algorithms for $b$-matchings \hbox{with~submodular~objectives}.

\textbf{Hypergraph Matching.} %
According to Hazan et~al.~\cite{approxresult}, the maximum $d$-set packing problem and, therefore, the matching problem on $d$-partite, $d$-uniform hypergraphs can be poorly approximated, and there is no approximation within a factor of $\mathcal{O}(d/\log d)$.
In general, as proven by Håstad~\cite{10.1007/BF02392825}, the matching problem in non-uniform hypergraphs and the maximum independent set problem are NP-hard and there is no $n^{1-
            \epsilon}$ factor approximation unless $P=NP$. 
There is a polynomial $(k + 1 + \epsilon)/3$-approximation algorithm for $k$-set packing, and therefore the matching problem in $d$-uniform, $d$-partite hypergraphs proposed by  Cygan~\cite{cygan2013improved} using local search.
Furthermore, Fürer and Yu~\cite{06825fa1bec94beab36a77646ab35159} improved these results with respect to the run-time.
Dufosse et al.~\cite{dufosse2019effective} introduce several heuristics to reduce the complexity of the uniform problem by extending the well-known two Karp-Sipser~\cite{karp1981maximum} rules to hypergraphs.
Dufosse et al.~\cite{dufosse2019effective} present the idea of using Sinkhorn-Knopp algorithm~\cite{sinkhorn1967concerning} for the normalization of incident tensors as a third selection rule. 
They perform practical experiments, but are limited to only $d$-partite, $d$-uniform hypergraphs with uniform edge weights.
Anneg et al.~\cite{annegetalnonuniform} give an improved optimality bound for LP-relaxation for the non-uniform case. \hbox{This extends to $b$-matching.}
 For the weighted $k$-set packing problem Thiery~and~Ward~\cite{doi:10.1137/1.9781611977554.ch42} show improved approximation bounds of $1.786$ for $k=3$.
 Recently, Neuwohner~\cite{doi:10.1137/1.9781611977554.ch41} showed how to proof a threshold below of  $\frac{k}{2}$ by $\Omega(k)$. Both approaches are improvements on long-standing local search  approach presented by Berman~\cite{10.1007/3-540-44985-X_19} for maximum weight independent set in $d$-claw~free~graphs.

\subparagraph{Hypergraph $b$-Matching.}
The $b$-matching cardinality problem in hypergraphs has also no approximation scheme according to El Ouali and Jäger~\cite{hypergraphbmatchingINNP}, even if the degree of vertices is bounded. Similarly, El Ouali et al.~\cite{khypergraph} showed that in $k$-uniform hypergraphs  for the cardinality problem with $2\leq b\leq k/\log k$ there is no polynomial-time approximation within any ratio smaller than $\Omega(\frac{k}{b\log{k}})$.
For weighted $b$-matching on $k$-uniform hypergraphs Krysta~\cite{krysta2005greedy} gave a greedy $k+1$ approximation, while Parekh and Pritchard~\cite{parekh2015generalized} achieve a $(k-1+\frac{1}{k})$
approximation algorithm via linear programming. Koufogiannakis and Young~\cite{koufogiannakis2009distributed} developed a $k$-approximation in a distributed fashion for weighted $k$-uniform hypergraphs. We are not aware of any practical implementation of~those~algorithms.

\section{Hypergraph $b$-Matching Algorithm}
\begin{algorithm}[t]
    \algsetup{linenosize=\scriptsize}
    {\footnotesize %
    \begin{algorithmic}[1]
        \PROCEDURE{BMatching}{$H=(V,\HypergraphEdges,\omega)$}
        \STATE $\mathcal{K},\mathcal{M}_{exact}, \HypergraphEdges_{folded} \gets \mathrm{Reductions}(H)$
        \STATE $\mathcal{M}_{Kernel} \gets \mathrm{Greedy}(\mathcal{K})\, \mathtt{ or }\,  \mathrm{ILP}(\mathcal{K})$
        \STATE $\mathcal{M}_{post}\gets \mathrm{Unfold}(H,\HypergraphEdges_{folded},\mathcal{M}_{Kernel})$
        \RETURN$\mathcal{M}_{exact}\cup \mathcal{M}_{Kernel}\cup \mathcal{M}_{post}$
        \ENDPROCEDURE
    \end{algorithmic}}
    \caption{$b$-Matching Algorithm}
    \label{alg:overview}
\end{algorithm}

We now give an overview of our algorithm to solve the general weighted hypergraph \hbox{$b$-matching} problem. 
In Section~\ref{sec:opt:solution} we introduce the exact integer linear program for this problem.
Our approach shown in Algorithm~\ref{alg:overview} starts by using exact data reductions devised in Section~\ref{sec:exact_reductions} to reduce the instance size. 
The reduced instances can then be used as input to the exact solver (based on the ILP) or  our heuristic algorithm. 
Our heuristic algorithm  computes a good initial $b$-matching using a greedy strategy, see Section~\ref{sec:greedy}.
 In Section~\ref{sec:localsearch} we introduce a local search, that improves solution quality by swaps.
Once a solution is computed on the reduced instance we reconstruct it to a solution for the original instance by unfolding the reductions in reverse order.

\subsection{Optimal Solutions}\label{sec:opt:solution}
To solve the $b$-matching problem to optimality, either on the original or exactly reduced hypergraph, we use the following integer linear program:
\begin{align}\label{eq:ilp}
    \max \sum_{e\in \HypergraphEdges} x_e\cdot \omega(e)\quad s.t.\,  \forall v\in V\colon \sum_{e\in \edgesOf{v}} x_e\leq b(v)\quad  x_e\in \{0,1\}\quad \forall e\in \HypergraphEdges.
\end{align}
For every edge $e\in E$, the integer linear program has a variable $x_e$, which is set to $x_e=1$ iff the edge $e$ is part of the matching and zero otherwise. The maximization term is the sum of the weights of the selected edges. The main constraint restricts the number of selected edges to obey to the capacity at each vertex in the original hypergraph.
\subsection{Exact Reduction Rules}\label{sec:exact_reductions}
Only few reductions are known that can be used for the hypergraph matching problem~\cite{dufosse2019effective}. These data reductions are based on Karp-Sipser rules and are a) not applicable to weighted problems and b) not applicable to the more general $b$-matching problem in hypergraphs.
However, especially for large instances, applying exact data reductions is a very important technique to decrease the problem size.
In general, reductions allow the classification of edges as either
(1) part of a solution,
(2) non-solution edges, or
(3) deferred, \ie the decision for this edge depends on additional information about neighboring edges that will be obtained later.
We denote by~$\mathcal{K}$ the resulting \emph{reduced hypergraph}, where no reduction rule applies anymore. 
In the following, we introduce a large set of new reductions for the weighted~$b$-matching~problem.

We now propose  our first data reduction rule, that is~the removal of abundant vertices. Intuitively, if the degree of a vertex is smaller or equal to its capacity, the vertex can be removed.
A vertex $v\in V$ is considered \emph{abundant} if its capacity $b(v)$ is equal to or exceeds its degree of edges.
\begin{reduction}[Abundant Vertices (AV)] \label{red:av}

    Any abundant vertex can be removed from the hypergraph. Moreover, any edge that becomes empty in the process can be included in~an~optimal~solution.
\end{reduction}
\begin{proof}
    An abundant vertex $v$ can be removed from the hypergraph as $v$ does not restrict the selection problem, that is, all incident edges of $v$ could in principle be contained in an optimum matching as the capacity is larger than the number of adjacent edges. 
    Thus, we can  remove $v$ from the hypergraph. If there is an edge~$e\in\HypergraphEdges$, only containing~$v$, it is part of an optimal solution, since it cannot be blocked at any other vertex.\hfill
\end{proof}
\begin{figure}[t]
    \centering
    \includegraphics[page=1]{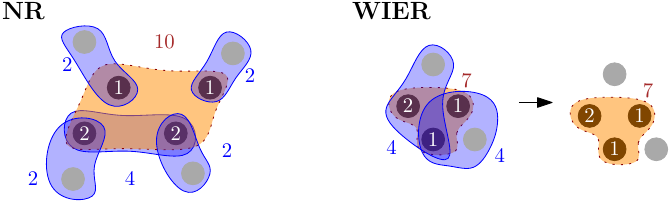}
    \caption{Examples for the first two reductions.  \textbf{Neighborhood Removal (NR):} The weight of the \dotuline{orange} edge dominates the sum of the $b(v)$-th heaviest weights per vertex (white), in this case the blue ones($10\geq 8$). There exist an optimal solution that contains the orange edge. \textbf{Weighted Isolated Edge Removal (WIER):} All edges form a clique and have a common vertex with capacity 1 each. The \dotuline{orange} edge is part of an optimal matching, because it has the highest weight of the clique.
    }
    \label{fig:nr:wier}
    
\end{figure}

\begin{reduction}[Neighborhood Removal (NR)]\label{red:nr}
    An edge $e\in E$ is in an optimal matching $\mathcal{M}_{opt}$ if~$e$ has a higher weight than the total sum of weights of the $b(v)$-th heaviest edge (excluding $e$) in each of its vertices $v\colon$
    $\omega(e)\geq \sum_{v\in \nodesOf{e}} \mathrm{nmax}(\hat{\omega}(E(v)\setminus \{e\}),b(v))$.
\end{reduction}

\begin{proof}
    Let $\mathcal{M}_{opt}$ be an optimal solution and assume $e\not\in \mathcal{M}_{opt}$.
    For each $v\in\nodesOf{e}$ with $b(v)=\lvert \mathcal{M}_{opt}(v)\rvert$ (conflicting vertices),
    we remove the lightest incident edge that is in the solution. Call this matching $\mathcal{M}'$. It follows for all vertices $v\in \nodesOf{e}\colon \lvert \mathcal{M}'(v)\rvert <b(v)$. This implies that $\mathcal{M}''=\mathcal{M}'\cup\{e\}$ is also a valid matching. 
    We now show $\omega(\mathcal{M}'')\geq\omega(\mathcal{M}_{opt})$.
    Let $e'_v$ be a lightest  edge removed from $\mathcal{M}_{opt}$ at $v$.
    Its weight $\omega(e'_v)$  contributing to $\mathcal{M}_{opt}$ is smaller or equal to the $b(v)$ heaviest edge incident to $v$, since $b(v)$ edges have been in $\mathcal{M}_{opt}$.
    Thus, ${\omega(e'_v)\leq \mathrm{nmax}(\hat{\omega}(\edgesOf{v}\setminus \{e\}),b(v))}$.
    The weight of all edges removed from $\mathcal{M}_{opt}$ is smaller or equal to $\omega(e)$, if the equation holds. This yields $\omega(\mathcal{M}'')\geq \omega(\mathcal{M}_{opt})$.~ \hfill
\end{proof}
Figure~\ref{fig:nr:wier} (NR) shows an example of Reduction~\ref{red:nr} and in Algorithm~\ref{alg:neighborhood_removal} a naive algorithm is presented.
 We iterate over each  vertex and check up to $b(v)$ incident edges. For each edge~$e$, we calculate the sum of weights it needs to dominate and break early if the condition cannot be satisfied anymore. If we find a candidate, we can include it as part of an optimal matching and update the hypergraph and capacity accordingly.
The time complexity for this algorithm is $\mathcal{O}(\min(n\beta,m)\Delta_E + n\Delta_V\log{\Delta_V})$. At each vertex, we have to check up to $\beta$ edges, and using a map for skipping already checked edges, we have in total up to $\mathcal{O}(\min(n\beta,m))$ candidates. The $\mathrm{nmax}$ operation is implementable in $\mathcal{O}(1)$ if the edge vector in each vertex is sorted. This initial sorting requires $\mathcal{O}(n\Delta_V\log{\Delta_V})$ steps. Note that we can find multiple reductions in the same pass. %

\newlength\XFigwidth\XFigwidth42mm

\begin{reduction}[Weighted Isolated Edge Removal (WIER)]\label{red:wie}
    Let $e\in E$ be an edge that has the heaviest weight of its neighbors: $\omega(e)\geq \max_{f\in \mathcal{N}(e)}\omega(f)$.
    If for all $ f,g\in \mathcal{N}(e)$ there exists a common vertex $v \in f\cap g$ with  capacity $b(v)=1$ then $e$ is part of an optimal solution.
\end{reduction}

\begin{proof}
    Because $e$ and all its adjacent edges contain at least one vertex with  capacity $1$, we can select at most one edge in $\mathcal{N}(e)$ while excluding all other edges in this neighborhood. An optimal solution $\mathcal{M}_{opt}$ must contain at least one edge of $\mathcal{N}(e)$, otherwise $e$ can directly be added yielding a heavier matching which is a contradiction to the optimality of the matching. %
    Given any optimal solution $\mathcal{M}_{opt}$ containing a neighbor $f$ then $\mathcal{M}_{opt}\setminus \{f\}\cup \{e\}$ is also optimal since  $\omega(e)\geq \max_{f\in \mathcal{N}(e)}\omega(f)$.\hfill
\end{proof}

\begin{figure}[t]
    \centering
    \includegraphics[page=4]{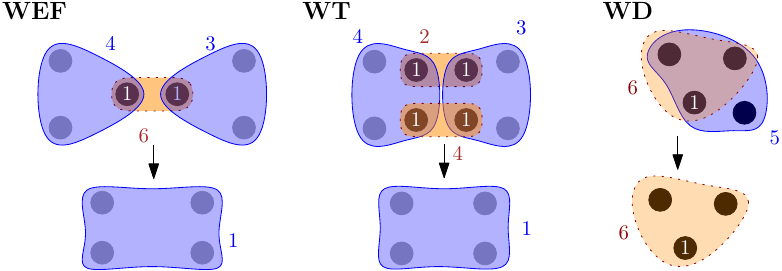}
    \caption{Examples for three reductions.
        \textbf{Weighted Edge Folding (WEF):}  The \dotuline{orange} edge has exactly two non-adjacent neighbors (blue), that it dominates one by one, but not in total. The three edges can be folded and later be decided~on.
        \textbf{Weighted Twin (WT):} The \dotuline{orange} edges have exactly two independent non-adjacent neighbors (blue), that they dominate one by one, but not in total. The four edges can be folded and later be~decided~on. 
        The Weighted Edge Folding is directly applied.
        \textbf{Weighted Domination (WD):} The \dotuline{orange} edge is a subset of the blue edge, has a higher weight and they share a common vertex with capacity~$1$.}
    \label{fig:wef}
\end{figure}

A hypergraph, where this reduction is applicable, is depicted in Figure~\ref{fig:nr:wier} (WIER).
In  Algorithm~\ref{alg:wie}, we give a procedure to detect  isolated edges and apply the reduction efficiently in detail.
At each vertex we have to scan the heaviest edge, but only if it has not been scanned before. We simultaneously check whether it has maximum weight at each other vertex and collect its neighbors. After checking if the weight condition is satisfied, we mark each edge incident to the currently scanned edge also as scanned. Because all neighbors weigh less than the currently scanned edge, they   cannot be candidates themselves for this reduction at other vertices later.
For neighbors at capacity~1 vertices, we save the position we scanned this edge in a binary encoding. We add neighboring edges at higher capacity vertices in a vector $N_l$. Afterwards, we check if all neighbors are incident to a vertex with capacity $1$ (property $S$) and if all pairs have a (blocking) common vertex of capacity~$1$. We use the binary encoding to check, if all edges in $N_l$ are incident to a capacity $1$ vertex. This greatly reduces the number of checks required.
This is accomplished by a \textit{bitwise and} of the encodings in $N_b$. If the result is non-zero we have to do a detailed check.
The overall complexity of this algorithm is $\mathcal{O}(\min(m,n){\Delta_E}^2)$ because in the worst case, we would collect~$\Delta_E$ distinct neighbors at a  vertex with $b(v)=1$ that we have to check for a common vertex.

The previous data reductions work by removing vertices (respectively, edges) from the graph. The following reduction modifies the structure of the hypergraph and postpones some  decisions to a later point.

\begin{figure*}
  \centering
   \begin{minipage}[t]{0.485\linewidth}
\begin{algorithm}[H]
    \algsetup{linenosize=\footnotesize}
    {\footnotesize
        \begin{algorithmic}[1]
            \PROCEDURE{NR}{$H=(V,\HypergraphEdges,\omega),b$}
            \STATE $C\gets \emptyset$
            \FOR{$v\in V$}
            \STATE{$checked\gets 0$}
            \FOR{$e\in \edgesOf{v} $ ordered by weight desc}
            \STATE $checked \gets checked+1$
            \IF{$checked>b(v)$}
            \STATE{\textbf{break}}
            \ENDIF
            \IF{$e\in C$}
            \STATE \textbf{continue}
            \ENDIF
            \STATE $C\gets C\cup \{e\}$
            \STATE $\omega_d\gets 0$
            \FOR{$w\in \nodesOf{e}$}
            \STATE $\omega_d\gets \omega_d+ \mathrm{nmax}(\hat{\omega}\left(E(w)\setminus \{e\}\right),b(w))$
            \IF{$\omega_d>\omega(e)$}
            \STATE \textbf{break}
            \ENDIF
            \ENDFOR 
            \IF{$\omega_d\leq \omega(e)$}
            \STATE $b'(v):=\begin{cases}
                    b(v)-1 & \text{if}\,v\in\nodesOf{e} \\
                    b(v)   & \text{else}
                \end{cases}$
            \STATE $R \gets \{e\}\cup blockedEdges(e) $
            \STATE \textbf{yield} $(H\setminus R,b'), \{e\},\emptyset$
            \ENDIF
            \ENDFOR
            \ENDFOR 
            \ENDPROCEDURE
        \end{algorithmic}
    }\caption{Neighborhood Removal}%
    \label{alg:neighborhood_removal}
\end{algorithm}

\begin{algorithm}[H]
    \algsetup{linenosize=\footnotesize}
    {\footnotesize
        \begin{algorithmic}[1]
            \PROCEDURE{WIER}{$H=(V,\HypergraphEdges,\omega),b$}
            \STATE $C\gets \emptyset$
            \FOR{$v\in V$ with $b(v)=1$} %
            \STATE $e\gets \mathrm{argmax}_{e\in\edgesOf{v}}\omega(e)$ %
            \IF{$e\in C$} %
            \STATE \textbf{continue} %
            \ENDIF
            \STATE $N_b\gets \{\}$
            \STATE $N_l\gets \emptyset$
            \STATE $count \gets 0$
            \STATE $B \gets True$
            \FOR{$w\in \nodesOf{e}$} %
            \IF{$\omega(e)<\max_{f\in\edgesOf{w}}\omega(f)$} %
            \STATE $B\gets False$
            \STATE \textbf{break}
            \ENDIF
            \FOR{$f\in \edgesOf{w}$}
            \STATE $C\gets C\cup \{f\}$
            \ENDFOR
            \IF{$b(w)=1$} %
            \FOR{$f \in \edgesOf{w}$} %
            \STATE $N_b[f] = N_b[f]+2^{count}$
            \ENDFOR
            \STATE $count \gets count + 1$
            \ELSE 
            \STATE $N_l\gets N_l \cup \edgesOf{w}$ %
            \ENDIF
            \ENDFOR 
            \STATE $S\gets \forall f\in N_l\colon N_b[f]>0$\COMMENT{ Check $N_l\subseteq N_b$}
            \FOR{$f,g\in N_b$} %
            \IF{$N_b[g]\& N_b[f] = 0 $} %
            \IF{$\not\exists w\in \nodesOf{e}\cap\nodesOf{f}: b(w)=1$} %
            \STATE $B \gets False$
            \STATE \textbf{break}
            \ENDIF
            \ENDIF
            \ENDFOR
            \IF{$ S \land B$}%
            \STATE \textbf{yield} $(H\setminus\{N_b\},b), \{e\},\emptyset$
            \ELSE
            \STATE $C\gets C\cup \{e\}$\hfill\COMMENT{Exclude at future vertices}
            \ENDIF
            \ENDFOR
            \ENDPROCEDURE
        \end{algorithmic}}\caption{ Weighted Isolated Edge Removal} %
    \label{alg:wie}
\end{algorithm}
\end{minipage}
\hfill
\begin{minipage}[t]{0.485\linewidth}
    \begin{algorithm}[H]
        \algsetup{linenosize=\footnotesize}
        {\footnotesize
            \begin{algorithmic}[1]
                \PROCEDURE{WEF}{$H=(V,\HypergraphEdges,\omega),b$}
                \STATE $C\gets\emptyset$
                \FOR{$v \in V$ with $b(v)=1$ and $\lvert \edgesOf{v}\rvert=2$}
                \FOR{$e\in \edgesOf{v}$ with $\lvert \nodesOf{e}\rvert= 2$} %
                \IF{$e\in C$}
                \STATE \textbf{continue}
                \ENDIF
                \STATE $C\gets C\cup \{e\}$
                \STATE $c \gets True$
                \STATE $N \gets \emptyset$
                \FOR{$w\in \nodesOf{e}$} %
                \IF{$\lvert \edgesOf{w}\rvert > 2\lor b(w)>1$}
                \STATE $c\gets False$
                \ELSE
                \STATE $N \gets N \cup  \edgesOf{w}\setminus\{e\}$ %
                \ENDIF
                \ENDFOR
                \IF{$c \land N$ independent} %
                \IF{$\omega(N)>\omega(e)\land \max_{e_n\in N}\omega(e_n)\leq \omega(e)$} %
                \STATE $\nodesOf{e_n}\gets \bigcup_{e\in N}\nodesOf{e}$
                \STATE $\omega'\gets \omega$
                \STATE $\omega'(e_n)\gets \omega(N)-\omega(e)$ %
                \STATE $\HypergraphEdges'  \gets \HypergraphEdges \setminus (N\cup \{e\}) \cup \{e_n\}$%
                \STATE \textbf{yield} $((V,\HypergraphEdges',\omega'),b), \emptyset,\{e\}$
                \ENDIF
                \ENDIF
                \ENDFOR
                \ENDFOR
                \ENDPROCEDURE
            \end{algorithmic}}\caption{Weighted Edge Folding}%
        \label{alg:wef}
    \end{algorithm}
    \vspace{0.125cm}
    \begin{algorithm}[H]
        \algsetup{linenosize=\footnotesize}
                 {\footnotesize
                     \begin{algorithmic}[1]
                         \PROCEDURE{TwinReduction}{$H=(V,\HypergraphEdges,\omega),b$}
                         \STATE $C\gets\emptyset$
                         \FOR{$v \in V$ with $b(v)=1$ and $\lvert \edgesOf{v}\rvert=2$}
                         \STATE $X \gets\emptyset$
                         \FOR{$e\in \edgesOf{v}$}%
                         \IF{$e\in C$}
                         \STATE \textbf{continue}
                         \ENDIF
                         \STATE $C \gets C\cup \{e\}$
                         \STATE $candidate \gets True$
                         \STATE $neighbors \gets \emptyset$
                         \FOR{$w\in \nodesOf{e}$} %
                         \IF{$\lvert w\rvert > 2\lor b(w)>1$}
                         \STATE $candidate\gets False$
                         \ELSE
                         \STATE $neighbors \gets neighbors \cup \edgesOf{w}\setminus\{e\}$%
                         \ENDIF
                         \ENDFOR
                         \IF{$candidate$}
                         \STATE $X\gets X\cup \{(e,neighbors)\}$
                         \ENDIF
                         \ENDFOR
                         \ENDFOR        
                         \FOR{$\exists N,e_1\neq e_2:(e_1,N),(e_2,N)\in X$} %
                         \IF{$N$ is independent} %
                         \STATE $\nodesOf{e_n}\gets \bigcup_{e\in N}\nodesOf{e}$
                         \STATE $\omega'\gets \omega$
                         \STATE $\omega'(e_n)\gets \omega(N)-\omega(\{e_1,e_2\})$
                         \STATE $\HypergraphEdges'  \gets (\HypergraphEdges \setminus \{e_1,e_2\}) \cup \{e_n\}$
                         \STATE $H'=(V,\HypergraphEdges',\omega')$
                         \IF{$\omega'(e')\geq \omega(N)$}
                         \STATE \textbf{yield} $\mathrm{NR}(H',b)$
                         \ELSIF{{$\omega'(e') > \omega(N)-\min_{e\in N}\omega(e)$}}
                         \STATE \textbf{yield} $\mathrm{EdgeFolding}(H',b)$
                         \ENDIF
                         \ENDIF
                         \ENDFOR
                         \ENDPROCEDURE
                     \end{algorithmic}}\caption{Weighted Twin}%
                 \label{alg:twin_reduction}
             \end{algorithm}
    \end{minipage}
\end{figure*}

\vspace{-0.25cm}
\begin{reduction}[Weighted Edge Folding (WEF)]\label{red:wef} Let $e\in\HypergraphEdges$ be a  edge and $N=\mathcal{N}(e)\setminus \{e\}$ be the edges adjacent to $e$. Suppose the following holds:
    \begin{enumerate}
        \item \label{proof:directneighbor} Each edge in $N$ is linked to $e$ via a vertex $v$ with capacity $b(v)=1$,
        \item \label{proof:independence} $N$ is independent, that is the vertices in all distinct $f,g\in N$ are disjoint,%
        \item \label{proof:dominance} The Neighborhood Removal is not applicable as $\omega(N)>\omega(e)$, but it holds $\omega(e)> \omega(N)-\min_{f\in N}\omega(f)$ %
    \end{enumerate}
    then we can \enquote{fold} $e$ and $N$ into a new edge $e'$, inducing an altered hypergraph~$H'$. The  hypergraph~$H'$ contains a new edge~$e'$ instead of $N$ and $e$ with  $\omega(e'):=\omega(N)-\omega(e)$ and $\nodesOf{e'}= \bigcup_{f\in N}\nodesOf{f}$. Let $\mathcal{M}'_{opt}$ be the optimal solution for $H'$. The weight of the maximum $b$-matching in $H$ is $\omega(\mathcal{M}'_{opt})+\omega(e)$. If the matching $\mathcal{M}_{opt}'$ contains $e'$ then $N$ is contained in an optimal solution for $H$. Otherwise, $e$ is contained in a maximum~matching~in~$H$.
\end{reduction}
\begin{proof}
    We first show, that either $e$ or  all edges in $N$ are contained in a maximum $b$-matching $\mathcal{M}_{opt}$.
    Assumption~\ref{proof:independence} guarantees that all edges in $N$ are possible candidates for $\mathcal{M}_{opt}$. Assumption~\ref{proof:directneighbor} allows  either $e$ or any edge of $N$ to be in a matching. %
    Let $F=N\cap \mathcal{M}_{opt}$ be a part of an optimal solution, we show that $F=N$ or $e\in \mathcal{M}_{opt}$. 
    We now assume that $F$ is nonempty and a strict subset of $N$, that is~$F\subsetneq N$. Due to Assumption~\ref{proof:directneighbor} this implies $e\not\in \mathcal{M}_{opt}$. 
    Since $F$ is a strict subset of $N$ and we have  $\omega(e)> \omega(N)-\min_{f\in N}\{\omega(f)\}\geq \omega(F)$ (Assumption~\ref{proof:dominance}), we could swap it for $e$ in $\mathcal{M}_{opt}$ and gain a better result.  This is a contradiction to $\mathcal{M}_{opt}$ being optimal and $F\subsetneq N$ being a strict subset. 
    If none of the edges in $N$ are part of $\mathcal{M}_{opt}$ $e$ is free and  we can include~it in the matching. 
    
    The vertices of $e'$ in $H'$ correspond to those of $N$ in $H$. The vertices of $e$ in $H$ are only contained in $e'$ in $H'$ without further neighbors. Therefore, if $e'$ is not in $\mathcal{M}'_{opt}$ the edge~$e$ must be in an optimal solution for $H$ and otherwise $N$ is included in an optimal solution~for~$H$.
    
    The formula for the weight is correct by the following case distinction. When $e'$ is not contained in $\mathcal{M}'$,  $e$ is free in the corresponding matching $\mathcal{M}$ and must be included in the optimal matching for $H$.
     Otherwise the weight of $\mathcal{M}'$  contains $\omega(e')$ and thus the optimal solution in $H$ has weight $\omega(M')+\omega(e)=\omega(\mathcal{M}'\setminus \{e'\})+\omega(e')+\omega(e)=\omega(\mathcal{M}'\setminus \{e'\})+\omega(N)-\omega(e)+\omega(e)=\omega(\mathcal{M}'\setminus \{e'\})+\omega(N)$. \hfill
\end{proof}

Algorithm~\ref{alg:wef} finds edges with two adjacent edges to fold in order to reduce the problem size. Only edges of size two are considered for complexity reasons. The complexity of this algorithm is $\mathcal{O}(\min(m,n)\Delta_E)$, because we have to check for $\mathcal{O}(\min(m,n))$ candidates if the two neighbors are independent which requires $\mathcal{O}(\Delta_E)$ checks. Without constraining the edge size, it would be $\mathcal{O}(\min(m,n)\Delta_V\Delta_E)$ since we would collect more neighbors.  %
We collect the neighbors on the two vertices with capacity~$1$ and check if they are independent. If so, we merge the independent neighboring edges and replace the neighbors and $e$ by this merged~$e_n$ with a new weight.
Figure~\ref{fig:wef} shows a (sub-)hypergraph where this reduction is applicable.

The following data reduction groups non-adjacent edges with the same independent neighborhood together. Afterwards, we can directly check again if \nameref{red:nr} or \nameref{red:wef} applies.

\begin{reduction}[Weighted Twin (WT)] \label{red:wt} Suppose  $e_1,e_2\in\HypergraphEdges$ are non-adjacent. Let $L_i=\mathcal{N}(e_i)\setminus\{e_i\}$  be the set of neigboring edges that are linked with $e_i$ via a vertex with capacity of 1. Assume each set $L_i$ is independent, $L_1=L_2$ and  $\omega(\{e_1,e_2\})>\omega(L_1)-\min_{f\in L_1} \omega(f)$ holds.  Then, we can solve the problem on a modified hypergraph~$H'$  of $H$, replacing $e_1,e_2$ with an edge~$e'$ with $\omega(e')=\omega(e_1)+\omega(e_2)$ and $\nodesOf{e'}=\nodesOf{e_1}$.%
\end{reduction}

\begin{proof}  Since $L_1=L_2$ and all edges in $L_2$ are linked via a capacity $1$ vertex we do not need to include the vertices in $\nodesOf{e_2}$. Because any capacity constraint for an edge  in $L_2$ at a vertex in $\nodesOf{e_2}$  is also present at a vertex in $\nodesOf{e_1}$.
    If $\omega(e')>\omega(L_1)$ the new edge $e'$ is weighing more than all of its neighbors combined, satisfying the condition of the \nameref{red:nr}. Otherwise, $\omega(e')>\omega(L_1)-\min_{n\in L_1} \omega(n)$ still holds and the properties for the \nameref{red:wef} are satisfied. Indeed, the neighbors $L_1$ and $e_1$ are linked, $L_1$ is independent  and the weight inequality for Assumption~\ref{proof:dominance} hold.\hfill%
\end{proof}

Figure~\ref{fig:wef} (WT) shows two edges that can be folded in such a way.
An algorithm for detecting twins is listed in Algorithm~\ref{alg:twin_reduction}. The algorithm first identifies all possible candidates that have only degree two vertices. Afterwards, we identify twins and either apply the Neighborhood Removal and add them directly to the matching or apply the Edge Folding. In this case, they only dominate their neighborhood except for one edge, and  we can merge the edges and assign a new weight to the new combined edge. In order to quickly find identical neighborhoods, we have to sort the candidates by neighborhood size. Each independence check requires $\mathcal{O}(\Delta_E\Delta_V)$ checks. The overall complexity of this algorithm is $\mathcal{O}(m\Delta_E\Delta_V+m\log{m})$. The algorithm for \hbox{\nameref{red:wef}} and \nameref{red:wt} share common steps and could be combined.

The original domination proposed by Fomin et al.~\cite{fomin2009measure} for the maximum  independent set reasoned that a vertex~$v$ having neighbors~$N_v$ is superfluous if there is a vertex~$w$ with subset~neighbors~$N_w\subset N_v$.
We extend this idea to edges in a weighted~hypergraph.

\begin{reduction}[Weighted Domination\hspace{0.05cm}(WD)] \label{red:wd}
    Let $e,f\in\HypergraphEdges$ be two  edges with $\omega(e)\geq \omega(f)$. Suppose $\nodesOf{e}$ is a subset of $\nodesOf{f}$ and there is a vertex $v\in\nodesOf{e}\subseteq\nodesOf{f}$ with capacity $b(v)=1$. Then, the edge~$f$ can be removed from the hypergraph.
\end{reduction}

\begin{proof} Since there is one vertex $v$ with $b(v)=1$ in $\nodesOf{e}\cap\nodesOf{f}$ the edges $e$ and $f$ cannot be both in the maximum matching at the same time. Now assume an optimal solution $\mathcal{M}_{opt}$ containing $f$. Since $\nodesOf{e}$ is a subset of $\nodesOf{f}$ and it holds $\omega(e)\geq \omega(f)$, in any optimal solution $e$ can replace $f$. Thus, $f$ can be removed from the hypergraph.%
    \hfill
\end{proof}
\begin{wrapfigure}{r}{0.5\linewidth}
    \vspace{-0.75cm}
    \begin{minipage}{\linewidth}
    \begin{algorithm}[H]
       \algsetup{linenosize=\footnotesize}
        {\footnotesize
            \begin{algorithmic}[1]
                \PROCEDURE{WD}{$H=(V,\HypergraphEdges,\omega),b$}
                \STATE $C\gets \emptyset$
                \FOR{$v\in V$ with $b(v)=1$}
                \FOR{$e_{s}\in \edgesOf{v}$ ordered by weight desc.}%
                \IF{$e_s\in C$}
                \STATE \textbf{continue}
                \ENDIF
                \STATE $C\gets \{e_s\}$
                \STATE $D \gets\emptyset$
                \FOR{$e\in \edgesOf{v},\omega(e_s)\geq \omega(e)$}%
                \IF{$\lvert \nodesOf{e}\rvert \leq \lvert \nodesOf{e_{s}}\rvert\land \mathrm{hash}(e,e_s)$}
                \STATE $D \gets D \cup \{e\}$
                \STATE \textbf{break}%
                \ENDIF
                \ENDFOR
                \FOR{$w\in \nodesOf{e_{s}}$}%
                \STATE $D \gets \{c \in D\mid w\in \nodesOf{c}\}$%
                \ENDFOR
                \STATE{$\HypergraphEdges'\gets \HypergraphEdges\setminus D$}%
                \RETURN $((V,\HypergraphEdges',\omega),b), \emptyset,\emptyset$
                \ENDFOR
                \ENDFOR
                \ENDPROCEDURE
            \end{algorithmic}}\caption{Weighted Domination}%
        \label{alg:wd}
    \end{algorithm}
    \end{minipage}
    \end{wrapfigure}

Figure~\ref{fig:wef} (WD) visualizes an example of this reduction, and  in  Algorithm~\ref{alg:wd} we show an implementation in pseudocode for finding a Weighted Domination. We iterate over each vertex and its incident edges in descending weight. We check if the following edges satisfy the natural size constraint and, using a simple hash function, check for the subset criterion. After collecting all candidates, we check which of these candidates are  super sets and remove them from the graph. The complexity with the shown subset check is $\mathcal{O}(m(\Delta_E\Delta_V\log{\Delta_V}))$ if we do not have a hashing function. For each edge, we can collect at most $\Delta_V-1$ candidates. For each candidate, we have to check  up to $\Delta_E$ vertices of $e$ that they are indeed incident requiring $\log{\Delta_V}$ comparisons if the list is sorted.   
By multiplying the id of vertices contained in an edge, storing these results in a wide integer and using the modulo operator to check for division without remainder, we can reduce the time complexity to $\mathcal{O}(m(\Delta_E+\Delta_V))$ in the ideal case. This requires a growable wide integer resulting in larger memory cost, so using  a hash function, like  multiplying only the $k$ least significant bits of the ids, seems reasonable. If $B$ is the width of the wide integer and we want to use the $k$ least significant bits, we can hash all edges if ${\Delta_E< (\frac{B}{k}\log(2)-1)}$~holds.

\subsection{Initial Solutions}
\label{sec:greedy}
We now present our algorithms to compute initial solutions. 
Roughly speaking, we use a greedy algorithm that sorts the edges by a priority function and adds free edges in this order. 
Note that the most intuitive order of adding heavy edges first may yield poor results. This is because any edge may block a wide range of other edges from being added, for example, if the current edge has a large number of vertices. Thus, other priority functions are necessary.
\label{sec:priorityfunc}
The core idea of our algorithm is to assign each edge a positive priority value and then add edges greedily in descending order of their priority (edges with the highest priority are added first).
To overcome the problems of the weight priority function explained above, we scale the weight function with the number of their incident vertices and with the capacity at each vertex of the edge. This ensures that we select edges first that a) have a high weight, b) do not block a lot of other edges, and c) have  vertices with a lot of remaining capacity.  We scale by capacity $
    h_{cap}(e)      := \omega(e)\prod_{v\in \nodesOf{e}}b(v)$, inverse edge size $h_{pin}(e)      := \frac{\omega(e)}{\lvert\nodesOf{e}\rvert} $, the combination of both $   h_{pin,cap}(e)  := \frac{\omega(e)}{\lvert\nodesOf{e}\rvert}\prod_{v\in \nodesOf{e}}b(v)$ and finally the capacity and inverse vertex degree $    h_{scaled}(e)   :=\omega(e)\prod_{v\in \nodesOf{e}}\frac{b(v)}{\lvert \edgesOf{v}\rvert}$.
    This results in four algorithms \textsf{cap}, \textsf{pin}, \textsf{pin,cap} and \textsf{scaled}, defined by the general framework and the respective objective function.

\subsection{Local Search}
\begin{algorithm}[t]
    {\footnotesize
        \begin{algorithmic}[1]
            \PROCEDURE{OneTwoSwap}{$H=(V,\HypergraphEdges,\omega),\mathcal{M}$}
            \FOR{$c\in \mathcal{M}$}%
            \STATE $l \gets \emptyset$
            \FOR{$p\in \nodesOf{c}$}%
            \FOR{$e\in  \edgesOf{p} \setminus \mathcal{M}$}%
            \IF{$ blocked(e,\mathcal{M})\subseteq blocked(c,\mathcal{M})$}
            \STATE $l\gets l \cup\{e\} $
            \ENDIF
            \ENDFOR
            \ENDFOR
            \IF{$\lvert l\rvert>1$}
            \STATE $\Phi(x) := blocked(x,\mathcal{M}\setminus\{c\}\cup\{x\})$
            \IF{$\exists x,y \in l: \Phi(y) \cap \Phi(x) = \emptyset\newline \textbf{and } \omega(x)+\omega(y)> \omega(c)$ }%
            \STATE $\mathcal{M}\gets \mathcal{M}\setminus\{c\}\cup \{x, y\}$%
            \STATE $\mathcal{M}\gets \mathrm{maximize}(\mathcal{M})$%
            \ENDIF
            \ENDIF
            \ENDFOR
            \ENDPROCEDURE
            \PROCEDURE{ExhaustiveOneTwoSwap}{$H,\mathcal{M}$}
            \STATE \textbf{while} solution improved \textbf{do}
            \STATE \quad OneTwoSwap($H,\mathcal{M}$)
            \ENDPROCEDURE
            \PROCEDURE{ILS}{$H=(V,\HypergraphEdges,\omega),\mathcal{M}$}
            \STATE $\mathcal{M}_{\text{best}} \gets  \mathcal{M}$
            \WHILE{stopping criterion not met}
            \STATE $\mathcal{M}_* \gets \mathrm{Perturb}(H,\mathcal{M})$
            \STATE ExhaustiveOneTwoSwap($H,\mathcal{M}_*$)
            
            \STATE $\mathcal{P} \gets \frac{1}{(\omega(\mathcal{M}_{\text{best}})-\omega(\mathcal{M}_*))(\omega( \mathcal{M}) -\omega( \mathcal{M}_*))}$
            \IF{$\omega(\mathcal{M}_*)>\omega(\mathcal{M})$}
            \STATE $\mathcal{M}\gets \mathcal{M}_*$
            \ELSIF{$x\in \mathrm{U}(0,1)< \mathcal{P}$}
            \STATE $\mathcal{M}\gets \mathcal{M}_*$
            \ENDIF
            \IF{$\omega( \mathcal{M}) > \omega(\mathcal{M}_{\text{best}})$}
            \STATE $\mathcal{M}_{\text{best}}\gets \mathcal{M}$
            \ENDIF
            \ENDWHILE
            \STATE
            \Return $\mathcal{M}_{\text{best}}$
            \ENDPROCEDURE
        \end{algorithmic}}\caption{\nameref{sec:localsearch}}
    \label{alg:ils}
\end{algorithm}
\label{sec:localsearch}
\label{sec:heuristics_improvement}
We now give a brief overview over our local search algorithm based on swapping. More detailed pseudocodes for the algorithms can be found in Algorithm~\ref{alg:ils}. A swap consists of removing one edge $e$ in favor of two edges $s_1,s_2$ that become unblocked by the removal of $e$. A swap is feasible if the combined weight of $s_1,s_2$ is greater than the weight of $e$.  %

\subparagraph{Swapping Algorithm.} %
In our swapping algorithm, we are searching for a matched edge and two non-matched adjacent edges that are only blocked by the matched edge and  can be admitted to the matching simultaneously \hbox{without conflict.} %

\subparagraph{$(1,2)$-Swaps.} For each edge in the matching $e$, we first collect all neighboring edges that satisfy the condition of being only blocked by $e$. 
In a second step, we identify a pair of edges that can be added without conflict when $e$ is removed from the matching and also constitute an improvement.
If we find such an edge pair, we include them in the matching and remove the edge $e$ from~it.
After a successful swap we maximize, \ie add all free edges to the solution.

\subparagraph{Perturbation.} The swapping algorithm ends up in a local optimum if executed repeatedly.
Therefore, we propose to perturb the solution (forcing hyperedges into the solution) similarly to Andrade et al.~\cite{DBLP:journals/heuristics/AndradeRW12} who do this for the independent set problem. 
The geometric distribution and the number of unmatched edges forced in the solution are directly adopted from Andrade~et~al.~\cite{DBLP:journals/heuristics/AndradeRW12}.

\subparagraph{Iterated Local Search.} The whole process is driven by the iterated local search. As long as the stopping criterion, in our case, a number of unsuccessful searches is not met, the  current solution $\mathcal{M}$ gets perturbed and then exhaustively improved by $(1,2)$-swaps. We call the obtained solution $\mathcal{M}_*$. A better solution automatically becomes the new starting point for the next iteration.
Similarly to Andrade~et~al.~\cite{DBLP:journals/heuristics/AndradeRW12}, we allow with a low probability\footnote{$\dfrac{1}{(\omega(\mathcal{M}_{\text{best}})-\omega(\mathcal{M}_*))(\omega( \mathcal{M}) -\omega( \mathcal{M}_*))}$} a slightly worse solution to be accepted as starting point of our next local search. 

\section{Experimental Evaluation}
\label{sec:Experimental Evaluation}
\subparagraph{Methodology.}
We implemented our algorithms and data structures in \textsf{C++17}. %
We compiled our program and all competitors using \textsf{g++-12.1} with full optimization turned on (-O3 flag).
In our experiments, we have used  machines provided by a cluster, equipped with two $20$-core Intel~Xeon~Gold~6230  processors running at $2.10$ GHz and having a cache of 27.5 MB. Each machine was either equipped with $96$ or $192$ GB of main memory. 
In general, we perform experiments with uniform capacities of $1$, $3$, $5$ and random capacity.
Random in this context means, that we assign each vertex a capacity uniformly distributed between 1 and its degree.
The results on uniform capacity closely follow the results for random capacity.
 Thus, these plots can be found in Appendix~\ref{app:further}.
We run the experiments~10 times and take the arithmetic mean as result per instance. We use SCIP~\cite{BestuzhevaEtal2021OO}, one of the fastest opensource ILP-Solvers, as a black box solver for ILPs.
 Deterministic experiments were only executed once iff the results (size, weight) and not the time was measured. The experiments were scheduled in parallel up to the numbers of physical cores of the machine, and the number of cores used by SCIP concurrently was limited to one. Each experiment run has a memory budget of $60$ GB per instance and $140$~GB in total. %

For comparison, we are using performance profiles as proposed by Dolan and Mor{\'e}~\cite{performanceprofiles}. We plot which fraction of instances is solved by an algorithm to an objective value of at least $\tau \omega(\mathcal{M}^\star)$, $0< \tau\leq 1$ and $\mathcal{M}^\star$ being the best matching reported by all heuristics. Thus, having a fraction near $1.0$ for a high $\tau$ is considered a good performance because a high fraction of instances is then solved to near~optimum.
Similarly, we are using these profiles to compare execution duration of approaches. Here $\tau$ is greater than $1$; for each instance, the time is marked relative as multiplicative of the minimum time needed to solve the instance~exactly by any approach.

\subparagraph*{Instances.} We use a wide range of instances collected from various sources to evaluate our algorithms and to compare against state-of-the-art competitors.
 We use the $M_{HG}$ dataset  of \textit{hypergraph instances}, provided by Gottesbüren et al.~\cite{DBLP:journals/corr/abs-2303-17679}  containing \textit{general hypergraphs}.
It consists  of \numprint{488} instances for four different use cases of hypergraphs, including \numprint{18} for circuit design (Ispd98,\cite{circuitbenchmark}), \numprint{10} routability-driven placement (Dac2012,\cite{dacset}), \numprint{184} instances derived from general matrices from the Suite Sparse Collection (SPM,\cite{sparsecollection}) and \numprint{276} instances derived from SAT solving problems (SAT,\cite{satbenchmark}).
When comparing against state-of-the-art $b$-matching algorithms in graphs, we use the 10 \textit{graphs} from the Florida Sparse Matrix Collection by Davis and Hu~\cite{sparsecollection}, as proposed by Khan et al.~\cite{khanbmatching} who also propose the algorithm that we compare~against.

\subparagraph{Edge Weights \& Capacity.} For general experiments, we assigned weights uniformly at random to the edges between $1$ and $100$. 
If we use random capacity, we sample for each vertex the capacity uniformly at random between one and its  vertex~degree.
\subparagraph{Overview.} The section is organized as follows. First, we compare our greedy priority functions from Section~\ref{sec:greedy} on hypergraphs. 
Furthermore, we compare our results with those of the $b$-matching graph algorithm \textsf{bSuitor} by Khan~et~al.~\cite{khanbmatching}.
 Then we study the effectiveness of our exact data reductions on SCIP~\cite{BestuzhevaEtal2021OO} as state-of-the-art opensource ILP solver. 
  Finally, we benchmark  our local search and compare against the competitors by Dufosse~et~al.~\cite{dufosse2019effective} on $6$-partite, $6$-uniform~hypergraphs.

\subparagraph{Priority Functions/Initial Solutions.}\label{exp:weight}

\begin{figure}[t]
    \centering
    \begin{subfigure}[t]{0.49\textwidth}
        \displaypgf{figures/experiments/tr_greedy}{performance_plot__performance_profile_-1_all}
        \caption{Quality results for the different proposed priority functions on \numprint{488} hypergraphs.}
        \label{fig:exp:greedy:heuer:hypergraphs}
    \end{subfigure}\hfill
    \begin{subfigure}[t]{0.49\textwidth}
        \displaypgf{figures/experiments/tr_bsuitor}{performance_plot__performance_profile_-1_all}
        \caption{Performance profile for  $b(v)=\mathrm{rnd}$ on graph instances selected by Khan~et~al.~\cite{khanbmatching}.} %
        \label{fig:exp:bsuitor:greedy}
    \end{subfigure}
    \caption{\vspace*{-10pt}}
\end{figure}

The priority functions for this experiment are defined in Section \ref{sec:greedy}. We test them on the general hypergraph data set. In Figure \ref{fig:exp:greedy:heuer:hypergraphs}, the results are shown for random capacity. For reference, we include the simple  greedy \textsf{weight} function without any scaling as a baseline. For uniform capacity of $1$, the \textsf{pin,cap} and \textsf{pin} compute the same best result since their objective function  for this capacity is the same. 
Our other two proposals are nearly identical to the simple greedy \textsf{weight} algorithm. For  a higher uniform capacity of $3$ or $5$ and random capacities, the quality of all of our proposals except for the \textsf{pin} is worse than the \textsf{weight} function.
The \textsf{pin} function finds the best solution in around 70 to 80 \% of the cases and requires a maximum $\tau=0.9$ across all capacities. %
Since we only need to precompute the values of \textsf{pin} once,
the running time of \textsf{pin} only slightly exceeds the running time of the simple \textsf{weight} algorithm, on average by 0.4\% of the \textsf{weight} function, which we consider neglectable.

We \emph{conclude} that \textsf{pin} is a natural good choice to compute initial solutions on general hypergraphs. The other proposals are not viable since they yield worse results than the greedy \textsf{weight} algorithm.
\subparagraph{Graph $b$-Matching.} In order to show the versatility of our approaches, we also tested them on normal graphs and compared our results to those by Khan~et~al.~\cite{khanbmatching}. We compiled and linked their program into our benchmark suite. The results of this first experiment are shown in Figure~\ref{fig:exp:bsuitor:greedy}. The competitor \textsf{bsuitor} by Khan~et~al.~\cite{khanbmatching} is run in comparison to our approaches introduced in Section~\ref{sec:greedy}.  Because of the fixed edge size of two in normal graphs and the capacity being static in all but one experiments, all but one function from Section~\ref{sec:greedy} report the same results. Only the \textsf{scaled} approach differs by taking the number of incident edges into account and yielding a better result. We can report an improvement of up to 2 \% over \textsf{bsuitor} on random capacity and up to 7 \% on capacity $1$. 
The running time of \textsf{bsuitor} on a single core in sequential can not be matched by our algorithms on graphs, these are on average 3.19 times slower. This is however expected since our data structure supports hypergraphs and that introduces an additional overhead when only~considering~graphs.
\subparagraph{Reductions and Speedup.}\label{sec:speedup:reductions}

\begin{figure}
    \centering
    \includegraphics{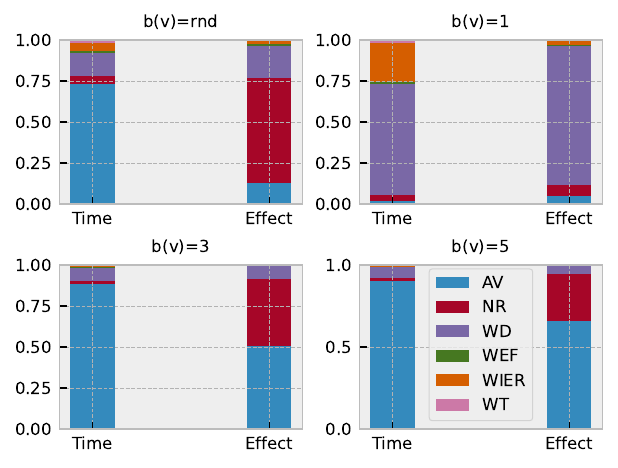}
    \caption{Relative running time and effect of reductions.}
    \label{fig:rel:runtime:eff}
\end{figure}

In this experiment, we investigate how well our reductions can speedup solving $b$-matching problems with SCIP~\cite{BestuzhevaEtal2021OO} as  an open source black-box solver.
As some of the data reductions can be expensive, we restrict some parameters in the search.  The parameters we use can be found in Table~\ref{tab:constraines:reductions}.  We empirically chose the parameters to balance between the reduction run time and success rate.
 The most sensitive tuning parameters were the ones for the \nameref{alg:wd}, since allowing bigger edges causes more checks to be needed. We apply our search algorithms for the reductions up to ten times and pass the resulting core problem to SCIP. We then compare its  total running time to the time it takes to solve the whole (hyper-)graphs without applying reductions. The running time of the program with reductions includes the time to find and apply the data reductions. %
We set a timeout for the SCIP computations of \numprint{1800} seconds and test it on uniform capacities of $1$, $3$ and $5$, as well as on random capacity. We ran this experiment once to determine which instances are solvable in the given time with any of these capacities and repeated the experiment on the solvable subset of $395$ hypergraphs~ten~times.
\\
Only instances on which \textsf{SCIP} returns the optimum within the time limit are considered in the time performance profile for random capacity shown in Figure~\ref{fig:exp:time}, similar results were obtained for the static capacities. %
The performance profiles show that there are some instances that are strongly affected by the reductions. Some instances are so small that the time to search for reductions is bigger than the solving itself. Overall, $80 \%$ of instances benefit from running the reductions search and $40 \%$ have a speedup of at least factor $2$.
In Table~\ref{tb:effect} we report the average and geometric speedup sorted by hypergraph class. Our reductions achieve the best average speed up on the \textrm{dac2012} instances of nearly $7$ for random capacity. These instances average edge count is halved by our reductions.  The least speed up is observed on the ISPD98 instances for uniform capacities. 
\\
A plot of the removal effectiveness is shown in Figure~\ref{fig:exp:removal_effectiveness}. On the $y$-axis, we display the relative edge count of the hypergraphs after applying our reductions. Similarly, on the $x$-axis, we plot the relative vertex count. Different shapes signify different capacities and colors are used for the different classes of hypergraphs. The majority of instances are located above the diagonal, meaning that the nodes are more reduced than the edges. There are two major clusters, one of not reducible instances and one for nearly completely reducible instances.
\\
Figure~\ref{fig:rel:runtime:eff} shows the relative running time of the reductions and effect on all 488 instances. The relative effect is computed on how many edges are included, excluded or deferred by the respective reductions.
For all capacities the Abundant Vertices, Neighborhood Removal and Weighted Domination are the most significant reductions and take up the most time.
The impact of the other reductions (Weighted Edge Folding, Weighted Isolated Edge Removal and Weighted Twin) is limited, but also the share of running time is small. 
For random assigned capacity the Neighborhood Removal reduction has the biggest impact, having over 50 \% of the effected edges, while taking only a small fraction of running time. 
The Weighted Domination contributes the most affected edges for uniform capacity of $1$, but also takes the most of the running time. 
The Weighted Isolated Edge Removal finds many candidates at capacity of $1$ and uses around one quarter of the running time, finding only few effected edges. 
For higher or random capacity the Abundant Vertices reduction requires most of the running time, but also contributes more than half of the effected edges in the static capacity case.

\begin{figure}
    \centering
    \begin{subfigure}[t]{0.49\textwidth}
        \displaypgf{figures/experiments/tr_reductions_greedy}{eff_rem__removal_plot}
        \caption{ Relative removal effectiveness. Some instances can be completely reduced, while many are not reduced at all.
        }
        \label{fig:exp:removal_effectiveness}
    \end{subfigure}\hfill
    \begin{subfigure}[t]{0.49\textwidth}
        \displaypgf{figures/experiments/tr_scip5}{time_performance_plot__performance_profile_-1_all}
        \caption{ Impact of reductions on optimum ILP solver \textsf{SCIP} on a subset of 395 hypegraphs. }
        \label{fig:exp:time}
        
    \end{subfigure}
    \caption{}
\end{figure}

\subparagraph{Local Search.}\label{sec:exp:local:search}
The iterated local search (\textsf{ILS}) approach from Section~\ref{sec:localsearch} is the main focus of this experiment. The \textsf{ILS} has only a threshold parameter $k$ determining after how many fruitless perturbations and local searches, the search is canceled. The best result so far is returned. In our experiment, we used the best solution obtained by \textsf{pin} function and tried to improve it. We chose $k=15$, as it showed a good balancing between running time and improvement on this collection of hypergraphs.
We report an improvement of up to eight percent at max over all capacities. Running time is around ten times on average longer than computing the initial solution. This is however not surprising as the initial algorithm only sorts by the priority function and adds edges greedily. The performance plot for this experiment is shown~in~Figure~\ref{fig:exp:impact:ilp:heuer:hypergraphs}. For reference, we include $k=50$ in the performance profile, which has an average running time tenfold in comparison to $k=15$. Some instances are only improved by a negligible amount, while the best instances are improved by $8\%$. The average improvement is around $3 \%$. %
\begin{figure}[t]\centering
    \begin{subfigure}[t]{0.49\textwidth}
        \displaypgf{figures/experiments/tr_ils_kss}{comb_both_performance_plot__performance_profile_1_all}
        \caption{ Solution quality of \textsf{Reductions+ILS}, \textsf{ksmd}, \textsf{kss} by Dufosse~et~al.~\cite{dufosse2019effective} and combination with \textsf{ILS}.} %
        \label{fig:exp:kss}
    \end{subfigure}\hfill
    \begin{subfigure}[t]{0.49\textwidth}
        \displaypgf{figures/experiments/tr_ils_rep}{performance_plot__performance_profile_-1_all}
        \caption{Iterated Local Search improves the solution quality by up to $8 \%$.
        } %
        \label{fig:exp:impact:ilp:heuer:hypergraphs}
    \end{subfigure}
    \caption{}
\end{figure}

\subparagraph{$1$-Matching on $d$-partite, $d$-uniform Hypergraphs.}\label{sec:exp:dufosse}
This experiment compares  and combines our algorithm with those developed by Dufosse~et~al.~\cite{dufosse2019effective}  on a special class of hypergraphs with uniform edge weight and uniform structure. They implemented \textsf{kss} and \textsf{ksmd}, which employ two configurations of the Karp-Sipser approach. We are using \textsf{kss} with $20$ scaling repetitions as proposed by the authors. We built and linked both algorithms into our benchmark program to ensure equal compile flags and settings. Figure~\ref{fig:exp:kss} shows  the performance profile for solution quality for $k=1500$, when combining \textsf{ILS} with their approaches or our approaches as initial solution. The number of retries needed to see an improvement is significantly higher than the one for the non-uniform problem. This is due to the uniform structure of the problem, we need more tries in the perturbation phase to find an improvement. %
For a low $k=50$, we can only report minimal improvements while running time stays closely to the base approaches.
Our \textsf{Reductions+ILS1500}, which combines the reductions, a greedy initial matching by \textsf{pin}, and iterated local search with $k=1500$ approach has a better solution quality than the \textsf{kss} and \textsf{ksmd} approaches.   Therefore, we want to combine both approaches. 

Our data reductions can not be combined with \textsf{kss} and \textsf{ksmd} since these algorithms require uniform edge size and our data reductions do not ensure this.
We can considerably improve the quality of \textsf{kss} and \textsf{ksmd} by using our \textsf{ILS} as postprocessing step. In case of \textsf{ksmd}, we can report a quality improvement of more than 30\% on around $8 \%$ of instances, even surpassing the results of the \textsf{Reductions+ILS1500} approach. Setting $k=1500$ unsuccessful tries can result on a few instances in a hundredfold running time while consistently yielding better quality. The improvement for \textsf{kss} is more pronounced than for \textsf{ksmd}, which has a specific reduction rule for this kind of hypergraph.
Improved quality comes at the expense of running time. The \textsf{ksmd} is the fastest approach and ten times faster than the \textsf{kss} approach. We need tenfold the time to compute the solution than \textsf{kss} on average.  Both, \textsf{ksmd} and \textsf{kss} are deterministic and thus can not be repeated multiple times to get a  fairer comparison. 
In conclusion, the choice of $k$ is sensitive for determining quality and running time and is dependent of the structure of the problem. Uniform weight problems require more tries than its non-uniform~counterparts.

\section{Conclusion}%
We developed a scalable algorithm for the general weighted $b$-matching problem in hypergraphs, utilizing novel data reductions. Our reductions can identify and incorporate optimum edges into a preliminary solution, reduce the problem size by combining the decision for multiple edges or, lastly, eliminate non-optimal edges from the input. %
We engineered new heuristic initial solution algorithms in a greedy framework and a local search framework to improve solutions.
 Experiments show that our data reductions scale well to large instances and accelerates state-of-the-art black-box solver. 
 The new initial heuristic solutions by a greedy framework are up to 10 \% better on general hypergraphs.
 On graphs these approaches also yield good results.
 The local search is able to improve solutions up to 30 \% over recent results by Dufosse~et~al.~\cite{dufosse2019effective} on some instances.
Given the good results, we will release our software as an open~source~project.
 Future work includes parallelization of our algorithms, research in nonlinear optimization objectives, improving local search techniques and application of our algorithms to \hbox{related~\NP{}-hard~problems}.

\bibliography{compactfixed.bib}

\begin{appendix}
    \section{Parameter of Reductions}
    {\footnotesize
    \begin{table}[H]
        \caption{Parameter constraints of the reductions. }
        \centering
        \begin{tabular}{llr}
            \toprule
            Reduction         & Constraint   & Default \\\midrule
            \nameref{red:nr}  & edge size    & 10      \\
            \nameref{red:wie} & edge size    & 8       \\
            \nameref{red:wie} & clique size  & 80      \\
            \nameref{red:wd}  & subedge size & 6       \\
            \nameref{red:wd}  & candidate    & 6       \\
            
            \nameref{red:wt}  & edge size    & 4       \\
            All               & iterations   & 10      \\\bottomrule
        \end{tabular}
        \label{tab:constraines:reductions}
        
    \end{table}
    }
    \section{Further Results}\label{app:further}
    \begin{table}[H]
        \caption{Removal effectiveness and speed up broken down for classes of hypergraphs. We report the average edge and vertex count per class. Only exactly solvable instances are considered.}
        \centering
        \hspace*{-1cm}
        \scalebox{0.85}{
            \begin{tabular}{rrrrrrrrr}
\hline
 &  & Instances & \# Edges (b.) & \# Edges & \# Vertices (b.) & \# Vertices & Avg Speedup & Geom. Speedup \\
Type & $b(v)$ &  &  &  &  &  &  &  \\
\hline
\multirow[t]{4}{*}{ISPD98} & 1 & 15 & \numprint{66149.73} & \numprint{63802.47} & \numprint{61454.80} & \numprint{51575.80} & \numprint{1.48} & \numprint{1.43} \\
 & 3 & 12 & \numprint{54146.83} & \numprint{34399.25} & \numprint{49819.42} & \numprint{21146.42} & \numprint{1.10} & \numprint{1.02} \\
 & 5 & 16 & \numprint{84121.75} & \numprint{27443.81} & \numprint{78844.19} & \numprint{15681.44} & \numprint{1.71} & \numprint{1.65} \\
 & rnd & 18 & \numprint{87241.94} & \numprint{57598.33} & \numprint{82194.44} & \numprint{26716.67} & \numprint{2.50} & \numprint{2.48} \\
\multirow[t]{4}{*}{SPM} & 1 & 76 & \numprint{90291.61} & \numprint{83653.75} & \numprint{66371.99} & \numprint{30211.46} & \numprint{2.60} & \numprint{1.74} \\
 & 3 & 85 & \numprint{106133.60} & \numprint{88644.16} & \numprint{93424.06} & \numprint{42218.27} & \numprint{1.64} & \numprint{1.21} \\
 & 5 & 95 & \numprint{121452.67} & \numprint{62287.86} & \numprint{109070.25} & \numprint{19574.86} & \numprint{2.86} & \numprint{1.57} \\
 & rnd & 149 & \numprint{101263.43} & \numprint{86516.18} & \numprint{91509.01} & \numprint{53173.65} & \numprint{1.52} & \numprint{1.29} \\
\multirow[t]{3}{*}{dac2012} & 3 & 9 & \numprint{870493.78} & \numprint{339460.11} & \numprint{883178.44} & \numprint{267180.00} & \numprint{3.14} & \numprint{2.96} \\
 & 5 & 10 & \numprint{912788.00} & \numprint{58781.30} & \numprint{924053.70} & \numprint{24115.70} & \numprint{7.12} & \numprint{6.66} \\
 & rnd & 10 & \numprint{912788.00} & \numprint{456581.90} & \numprint{924053.70} & \numprint{143152.90} & \numprint{6.98} & \numprint{6.85} \\
\multirow[t]{4}{*}{sat14} & 1 & 88 & \numprint{174182.83} & \numprint{171749.31} & \numprint{132746.41} & \numprint{121557.12} & \numprint{2.61} & \numprint{1.50} \\
 & 3 & 154 & \numprint{378729.73} & \numprint{210010.55} & \numprint{980506.36} & \numprint{83418.93} & \numprint{3.71} & \numprint{2.18} \\
 & 5 & 164 & \numprint{515168.57} & \numprint{226712.89} & \numprint{1010003.43} & \numprint{34002.97} & \numprint{5.77} & \numprint{3.00} \\
 & rnd & 155 & \numprint{281539.58} & \numprint{211308.96} & \numprint{302720.74} & \numprint{76921.45} & \numprint{2.38} & \numprint{2.05} \\
\multirow[t]{4}{*}{all} & 1 & 179 & \numprint{129511.16} & \numprint{125299.78} & \numprint{98590.93} & \numprint{76909.13} & \numprint{2.51} & \numprint{1.59} \\
 & 3 & 260 & \numprint{291653.62} & \numprint{166708.88} & \numprint{644174.86} & \numprint{73436.33} & \numprint{2.89} & \numprint{1.75} \\
 & 5 & 285 & \numprint{373682.38} & \numprint{154825.18} & \numprint{654400.99} & \numprint{27818.10} & \numprint{4.62} & \numprint{2.40} \\
 & rnd & 332 & \numprint{209111.81} & \numprint{154356.59} & \numprint{214688.53} & \numprint{65536.53} & \numprint{2.14} & \numprint{1.74} \\
\hline
\end{tabular}

        }
        \label{tb:effect}
    \end{table}
\begin{figure}
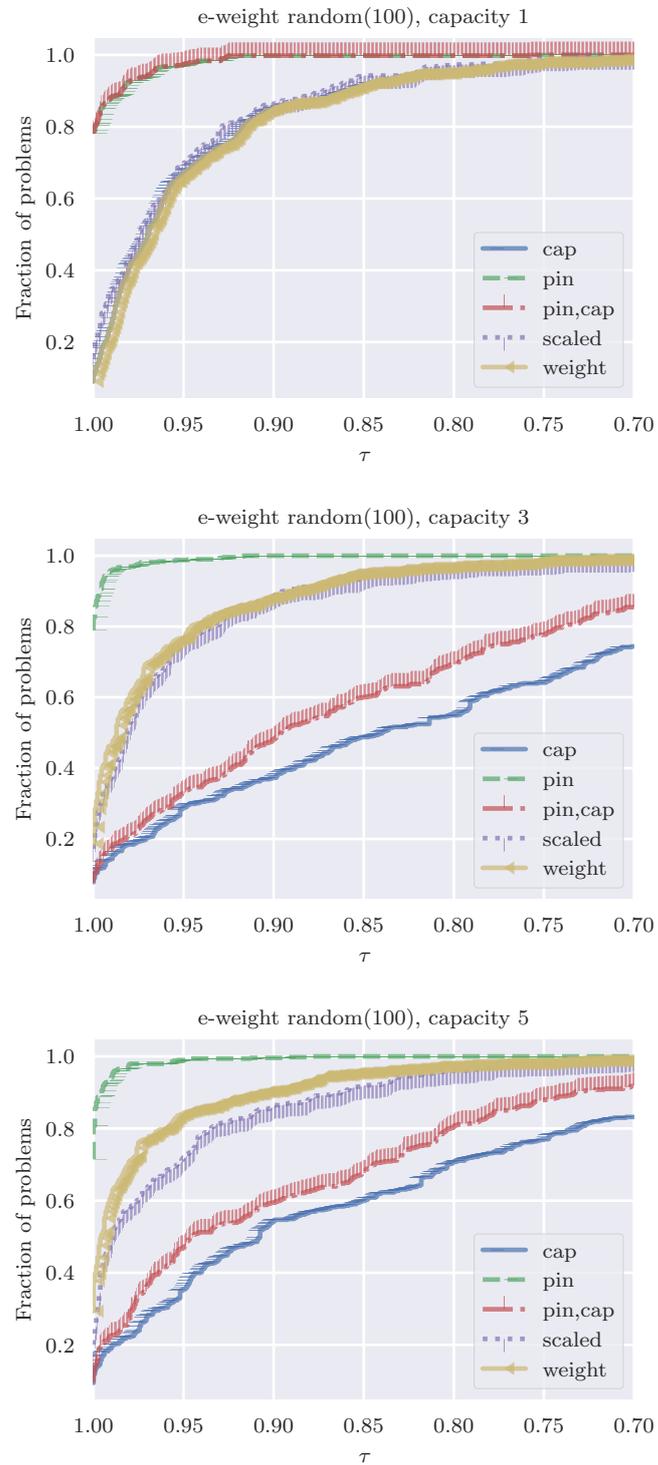

    \centering
    \displaypgf{figures/experiments/tr_greedy}{performance_plot__performance_profile_1_all}
    \displaypgf{figures/experiments/tr_greedy}{performance_plot__performance_profile_3_all}
    \displaypgf{figures/experiments/tr_greedy}{performance_plot__performance_profile_5_all}
    \caption{Quality results for the different proposed priority functions on 488 hypergraphs for constant $b(v)=\{1,3,5\}$.}
\end{figure}
    \begin{figure}
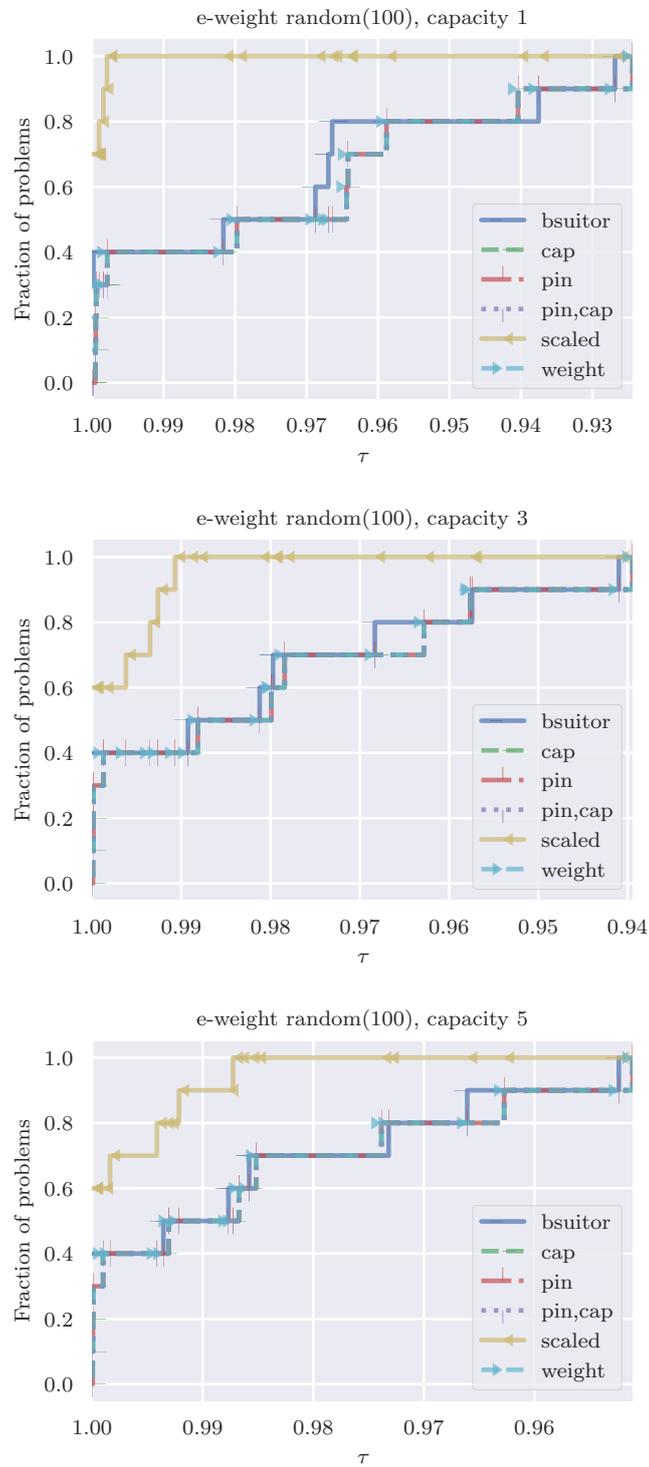

    \centering
        \displaypgf{figures/experiments/tr_bsuitor}{performance_plot__performance_profile_1_all}
        \displaypgf{figures/experiments/tr_bsuitor}{performance_plot__performance_profile_3_all}
        \displaypgf{figures/experiments/tr_bsuitor}{performance_plot__performance_profile_5_all}

        \caption{Performance profile for  $b(v)=\{1,3,5\}$ on graph instances selected by Khan~et~al.~\cite{khanbmatching}.}
    \end{figure}
   
    \begin{figure}
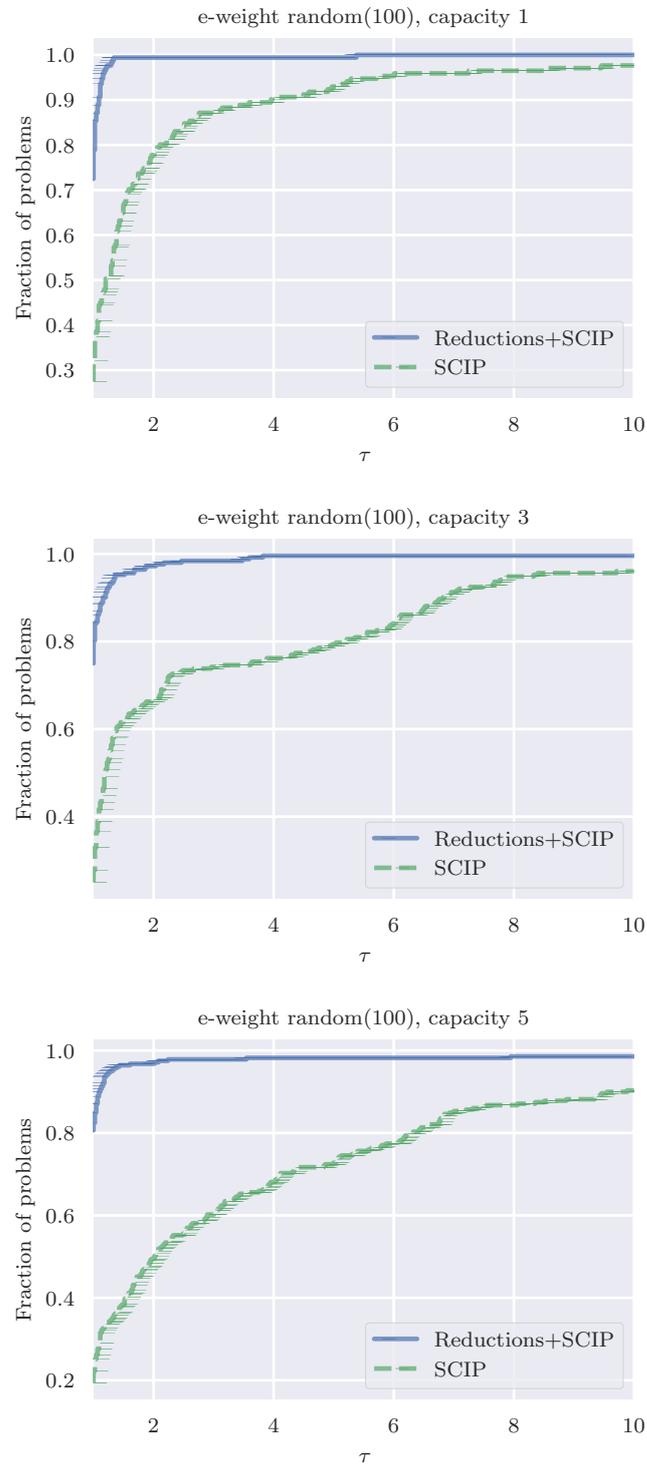

    \centering
        \displaypgf{figures/experiments/tr_scip5}{time_performance_plot__performance_profile_1_all}
        \displaypgf{figures/experiments/tr_scip5}{time_performance_plot__performance_profile_3_all}
        \displaypgf{figures/experiments/tr_scip5}{time_performance_plot__performance_profile_5_all}
        \caption{ Impact of reductions on optimum ILP solver \textsf{SCIP} on a subset of 395 hypegraphs for static capacities. }
    \end{figure}
    \begin{figure}
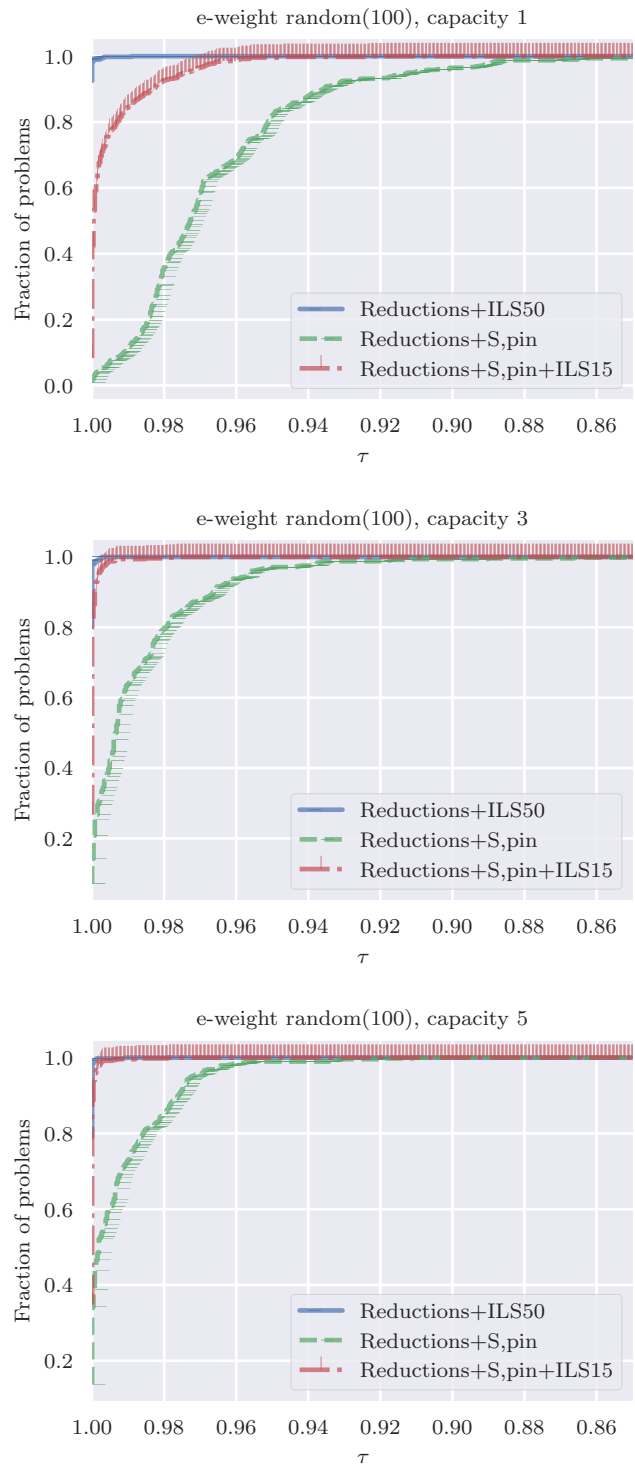

    \centering
        \displaypgf{figures/experiments/tr_ils_rep}{performance_plot__performance_profile_1_all}
        \displaypgf{figures/experiments/tr_ils_rep}{performance_plot__performance_profile_3_all}
        \displaypgf{figures/experiments/tr_ils_rep}{performance_plot__performance_profile_5_all}
        \caption{Further results for varying static capacity on hypergraphs.}
    \end{figure}

\end{appendix}
\end{document}